\documentclass[preprint,amsmath,amssymb,11pt,english,aps,showpacs,showkeys]{revtex4}
\usepackage{graphicx}
\usepackage{graphics}
\usepackage{amsmath}
\usepackage{dcolumn}
\usepackage{amssymb}
\usepackage{bm}

\begin{document}
\title{On the confinement of a quantum particle to a two-dimensional ring in systems described by the Dirac equation}
\author{K. Bakke}
\email{kbakke@fisica.ufpb.br}
\affiliation{Departamento de F\'isica, Universidade Federal da Para\'iba, Caixa Postal 5008, 58051-970, Jo\~ao Pessoa, PB, Brazil.}
\author{C. Furtado}
\affiliation{Departamento de F\'isica, Universidade Federal da Para\'iba, Caixa Postal 5008, 58051-970, Jo\~ao Pessoa, PB, Brazil.}
\email{furtado@fisica.ufpb.br}

\begin{abstract}
In this contribution, we propose a new model for studying the confinement of a spin-half particle to a two-dimensional quantum ring in systems described by the Dirac equation by introducing a new minimal coupling into the Dirac equation. We show that the introduction of this new minimal coupling into the Dirac equation yields a generalization of the two-dimensional model for a quantum ring proposed by Tan and Inkson [W.-C. Tan and J. C. Inkson, Semicond. Sci. Technol. {\bf11}, 1635 (1996)] for relativistic spin-half quantum particles.
\end{abstract}


\maketitle


Recent studies of the interaction between a relativistic spin-half particle with the harmonic oscillator potential has shown in the nonrelativistic limit of the Dirac equation the impossibility of recovering the harmonic oscillator Hamiltonian due to the presence of a quadratic potential \cite{do1,do2,do3,do4}, therefore it was introduced a minimal coupling into the Dirac equation in such a way that the Dirac equation remains a linear equation in both momenta and coordinates, and, in the nonrelativistic limit of Dirac equation, one can recover the Schr\"odinger equation for a harmonic oscillator. This new coupling introduced into the Dirac equation is called the Dirac oscillator \cite{do1}. In recent years, the Dirac oscillator has been used in studies of Ramsey-interferometry effect \cite{do4}, in quantum Hall effect \cite{do5}, and in the presence of an external magnetic field \cite{do6}.

However, the introduction of the Dirac oscillator does not allow us to make a complete study of the confinement of a relativistic quantum particle to a two-dimensional quantum ring. In the nonrelativistic context of the quantum mechanics, the confinement of quantum particles into a two-dimensional quantum ring has presented interesting results, such as, a nonparabolic spectrum of energy \cite{tan}, the arising of persistent currents due to the dependency of the energy levels the Berry's phase \cite{ring}, a parabolic spectrum of energy \cite{dot,dot2,ani,bf18}, and the arising of persistent currents due to the dependency of the energy levels the Aharonov-Casher quantum flux \cite{ring7,bf18}.

In this letter, based on the minimal coupling corresponding to the Dirac oscillator \cite{do1}, we propose the introduction of a new minimal coupling into the Dirac equation to study the confinement of a spin-half particle to a two-dimensional quantum ring in condensed matter systems described by the Dirac equation. Systems described by the Dirac equation are characterized by a linear dispersion with the velocity obeyed by the quasiparticles. For instance, near the Fermi points in graphene, the dispersion relation is linear with the momenta, where the Fermi velocity plays the hole of velocity of the light \cite{lc1,lc2}. We show that the introduction of this new minimal coupling into the Dirac equation yields a generalization of the two-dimensional model for a quantum ring proposed by Tan and Inkson \cite{tan} for relativistic quantum particles, and, in the nonrelativistic limit, we can obtain a nonparabolic and discrete spectrum of energy for a spin-half particle confined to two-dimensional quantum ring analogous to the Tan-Inkson model \cite{tan}. Furthermore, we show that this relativistic model allows us to discuss both the confinement of a spin-half particle to a quantum dot, and the interaction with a quantum antidot in systems described by the Dirac equation. We begin by by introducing a new minimal coupling into the Dirac equation. We also consider the presence of a magnetic flux in the center of the two-dimensional quantum ring, and solve the Dirac equation exactly. We show that the relativistic energy levels has a dependence of the magnetic quantum flux which gives rise to the arising of persistent currents in the two-dimensional quantum ring. In the following, we discuss the limit where this model describes the confinement of a spin-half particle to a quantum dot, and the interaction with a quantum antidot in systems described by the Dirac equation. At the end, we obtain the Dirac spinors for positive-energy solutions.

Let us begin by introducing a new minimal coupling into the Dirac equation. As we have discussed in the introduction, the impossibility of recovering the harmonic oscillator Hamiltonian in the nonrelativistic limit of the Dirac equation by adding a parabolic scalar potential into the Dirac equation \cite{do1,do2,do3,do4}, gave rise to the introduction a minimal coupling in such a way that the Dirac equation could remain a linear equation in both momenta and coordinates. This minimal coupling is known in the literature as the Dirac oscillator \cite{do1}, where it is given by
\begin{eqnarray}
\vec{p}\rightarrow\vec{p}-im\omega\rho\,\gamma^{0}\,\hat{\rho},
\label{1}
\end{eqnarray}
where $\hat{\rho}$ is an unit vector on the radial direction. In this way, based on the minimal coupling which gives rise to the Dirac oscillator, we introduce the following minimal coupling into the Dirac equation
\begin{eqnarray}
\vec{p}\rightarrow\vec{p}+i\left[\frac{\sqrt{2ma_{1}}}{\rho}+\sqrt{2ma_{2}}\,\rho\,\right]\gamma^{0}\,\hat{\rho},
\label{2}
\end{eqnarray} 
where $a_{1}$ and $a_{2}$ are control parameters. In the following, we show that the introduction of the minimal coupling (\ref{2}) gives rise to a model for studies of the confinement of a relativistic spin-half particle to a two-dimensional quantum ring yielding a generalization of the Tan-Inkson model for a two-dimensional quantum ring \cite{tan}. We also show that, by taking the parameter $a_{1}=0$, the relativistic energy levels correspond to both the energy levels of the Dirac oscillator, and the confinement of a relativistic spin-half particle to a quantum dot. Moreover, by taking the parameter $a_{2}=0$, we have the case where the relativistic spin-half particle interacts with a quantum antidot, and no bound states can be achieved.

Since the minimal coupling (\ref{2}) takes into account the cylindrical symmetry, we work the Dirac equation with curvilinear coordinates. In curvilinear coordinates, for instance, in cylindrical coordinates, the line element of the Minkowski spacetime is written in the form: $ds^{2}=-dt^{2}+d\rho^{2}+\rho^{2}d\varphi^{2}+dz^{2}$. Thus, by applying a coordinate transformation $\frac{\partial}{\partial x^{\mu}}=\frac{\partial \bar{x}^{\nu}}{\partial x^{\mu}}\,\frac{\partial}{\partial\bar{x}^{\nu}}$, and a unitary transformation on the wave function $\psi\left(x\right)=U\,\psi'\left(\bar{x}\right)$, the Dirac equation can be written in any orthogonal system in the following form \cite{schu}:
\begin{eqnarray}
i\,\gamma^{\mu}\,D_{\mu}\,\psi+\frac{i}{2}\,\sum_{k=1}^{3}\,\gamma^{k}\,\left[D_{k}\,\ln\left(\frac{h_{1}\,h_{2}\,h_{3}}{h_{k}}\right)\right]\psi=m\psi,
\label{2.1}
\end{eqnarray}
where $D_{\mu}=\frac{1}{h_{\mu}}\,\partial_{\mu}$ is the derivative of the corresponding coordinate system, and the parameters $h_{k}$ correspond to the scale factors of this coordinate system. For instance, in cylindrical coordinates, the scale factors are $h_{0}=1$, $h_{1}=1$, $h_{2}=\rho$, and $h_{3}=1$. In this way, the second term in (\ref{2.1}) gives rise to a term called the spinorial connection \cite{schu,bf10,b4,bbs,bbs2,weinberg}. The matrices $\gamma^{\mu}$ are the Dirac matrices given in the Minkowski spacetime \cite{bjd}, that is,
\begin{eqnarray}
\gamma^{0}=\hat{\beta}=\left(
\begin{array}{cc}
1 & 0 \\
0 & -1 \\
\end{array}\right);\,\,\,
\gamma^{i}=\hat{\beta}\,\hat{\alpha}^{i}=\left(
\begin{array}{cc}
 0 & \sigma^{i} \\
-\sigma^{i} & 0 \\
\end{array}\right);\,\,\,\Sigma^{i}=\left(
\begin{array}{cc}
\sigma^{i} & 0 \\
0 & \sigma^{i} \\	
\end{array}\right),
\label{2}
\end{eqnarray}
with $I$ being the $2\times2$ identity matrix, $\vec{\Sigma}$ being the spin vector, and $\sigma^{i}$ being the Pauli matrices. The Pauli matrices satisfy the relation $\left(\sigma^{i}\,\sigma^{j}+\sigma^{j}\,\sigma^{i}\right)=2\,\eta^{ij}$, where $\eta^{\mu\nu}=\mathrm{diag}\left(-\,+\,+\,+\right)$ is the Minkowski tensor.

Let us also consider the presence of a magnetic flux given by $\vec{A}=\frac{\phi}{2\pi\rho}\,\hat{e}_{\varphi}$, with $\phi$ being the magnetic flux on the $z$ direction, where the magnetic field is given by $\vec{B}=\phi\,\delta\left(x\right)\,\delta\left(y\right)\,\hat{z}$. Thus, by introducing the minimal coupling (\ref{2}) into the Dirac equation in the presence of the magnetic flux, the Dirac equation becomes
\begin{eqnarray}
i\frac{\partial\psi}{\partial t}=m\hat{\beta}\psi-i\hat{\alpha}^{1}\left[\frac{\partial}{\partial\rho}+\frac{1}{2\rho}-\hat{\beta}\frac{\sqrt{2ma_{1}}}{\rho}-\hat{\beta}\,\sqrt{2ma_{2}}\rho\right]\psi-i\frac{\hat{\alpha}^{2}}{\rho}\left[\frac{\partial}{\partial\varphi}-i\frac{\phi}{\phi_{0}}\right]\psi-i\hat{\alpha}^{3}\frac{\partial\psi}{\partial z},
\label{3}
\end{eqnarray}
where $q$ corresponds to the electric charge of the particle, and $\phi_{0}=2\pi/\left|q\right|$. The solution of the Dirac equation (\ref{3}) is given in the form:
\begin{eqnarray}
\psi=e^{-i\mathcal{E}t}\left(
\begin{array}{c}
\eta\\
\chi\\	
\end{array}\right),
\label{4}
\end{eqnarray}
where $\eta=\eta\left(\rho,\varphi,z\right)$ and $\chi=\chi\left(\rho,\varphi,z\right)$ are two-spinors. Thus, substituting (\ref{4}) into the Dirac equation (\ref{3}), we obtain two coupled equations for $\eta$ and $\chi$, where the first coupled equation is
\begin{eqnarray}
\left(\mathcal{E}-m\right)\eta=-i\sigma^{1}\left[\frac{\partial}{\partial\rho}+\frac{1}{2\rho}+\frac{\sqrt{2ma_{1}}}{\rho}+\,\sqrt{2ma_{2}}\,\rho\right]\chi-i\frac{\sigma^{2}}{\rho}\left[\frac{\partial}{\partial\varphi}-i\frac{\phi}{\phi_{0}}\right]\chi-i\sigma^{3}\frac{\partial\chi}{\partial z},
\label{5}
\end{eqnarray}
while the second coupled equation is
\begin{eqnarray}
\left(\mathcal{E}+m\right)\chi=-i\sigma^{1}\left[\frac{\partial}{\partial\rho}+\frac{1}{2\rho}-\frac{\sqrt{2ma_{1}}}{\rho}-\,\sqrt{2ma_{2}}\,\rho\right]\eta-i\frac{\sigma^{2}}{\rho}\left[\frac{\partial}{\partial\varphi}-i\frac{\phi}{\phi_{0}}\right]\eta-i\sigma^{3}\frac{\partial\eta}{\partial z},
\label{6}
\end{eqnarray}

By eliminating $\chi$ in (\ref{6}), and substituting in (\ref{5}), we obtain the following second order differential equation
\begin{eqnarray}
\left(\mathcal{E}^{2}-m^{2}\right)\eta&=&-\frac{\partial^{2}\eta}{\partial\rho^{2}}-\frac{1}{\rho}\,\frac{\partial\eta}{\partial\rho}+\frac{\eta}{4\rho^{2}}-\frac{\sqrt{2ma_{1}}}{\rho}\,\eta+\sqrt{2ma_{2}}\,\eta+i\frac{\sigma^{3}}{\rho^{2}}\,\frac{\partial\eta}{\partial\varphi}+\frac{2ma_{1}}{\rho^{2}}\,\eta\nonumber\\
&+&4m\sqrt{a_{1}a_{2}}\,\eta-2i\sigma^{3}\,\frac{\sqrt{2ma_{1}}}{\rho^{2}}\,\frac{\partial\eta}{\partial\varphi}-2\sigma^{3}\,\frac{\phi}{\phi_{0}}\frac{\sqrt{2ma_{1}}}{\rho^{2}}\,\eta+2ma_{2}\,\rho^{2}\,\eta\nonumber\\
[-2mm]\label{7}\\[-2mm]
&-&2i\sigma^{3}\,\sqrt{2ma_{2}}\,\frac{\partial\eta}{\partial\varphi}-2\frac{\phi}{\phi_{0}}\,\sqrt{2ma_{2}}\,\sigma^{3}\eta+\frac{\phi}{\phi_{0}}\,\frac{\sigma^{3}}{\rho^{2}}\,\eta-\frac{1}{\rho^{2}}\,\frac{\partial^{2}\eta}{\partial\varphi^{2}}\nonumber\\
&+&\left(\frac{\phi}{\phi_{0}\rho}\right)^{2}\,\eta+2i\frac{\phi}{\phi_{0}\rho^{2}}\,\frac{\partial\eta}{\partial\varphi}-\frac{\partial^{2}\eta}{\partial z^{2}}\nonumber,
\end{eqnarray}
where we have assumed that the spin of the quantum particle is polarized on the $z$-axis. We can see that $\eta$ is an eigenfunction of $\sigma^{3}$, whose eigenvalues are $s=\pm1$. Thus, we can write $\sigma^{3}\eta_{s}=\pm\eta_{s}=s\eta_{s}$, that is, $\sigma^{3}\,\eta_{+}=\eta_{+}$ and $\sigma^{3}\,\eta_{-}=-\eta_{-}$. We can also see that the operators $\hat{J}_{z}=-i\partial_{\varphi}+\frac{\sigma^{3}}{2}$ and $\hat{p}_{z}=-i\partial_{z}$ commute with the Hamiltonian of the right-hand side of (\ref{7}), then,  we can take the solutions of (\ref{7}) in the form:
\begin{eqnarray}
\eta_{s}=e^{i\left(l+\frac{1}{2}\right)\varphi}\,e^{ikz}\,\left(
\begin{array}{c}
R_{+}\left(\rho\right)\\
R_{-}\left(\rho\right)\\	
\end{array}\right),
\label{8}
\end{eqnarray}
where $l=0,\pm1,\pm2,\ldots$ and $k$ is a constant. Since we are interested in discussing a planar system, thus, we consider $k=0$. Substituting (\ref{8}) into (\ref{7}), we obtain two non-coupled radial equations for $R_{+}\left(\rho\right)$ and $R_{-}\left(\rho\right)$, therefore we write these equation in the following compact form:
\begin{eqnarray}
\left[\frac{d^{2}}{d\rho^{2}}+\frac{1}{\rho}\,\frac{d}{d\rho}-\frac{\tau_{s}^{2}}{\rho^{2}}-2ma_{2}\,\rho^{2}+\nu_{s}\right]\,R_{s}\left(\rho\right)=0,
\label{9}
\end{eqnarray}
where we have defined in Eq. (\ref{9}), the parameters:
\begin{eqnarray}
\tau_{s}&=&\zeta_{s}+s\sqrt{2ma_{1}};\nonumber\\
\zeta_{s}&=&l+\frac{1}{2}\left(1-s\right)-\frac{\phi}{\phi_{0}};\\ \label{10}
\nu_{s}&=&\mathcal{E}^{2}+m^{2}-2s\sqrt{2ma_{2}}\,\zeta_{s}-2\sqrt{2ma_{2}}-4m\sqrt{a_{1}a_{2}}.\nonumber
\end{eqnarray}

Before solving the second order differential equation (\ref{9}), we make a coordinate transformation given by $\mu=\sqrt{2ma_{2}}\,\rho^{2}$, and rewrite (\ref{9}) in the form:
\begin{eqnarray}
\left[\mu\frac{d^{2}}{d\mu^{2}}+\frac{d}{d\mu}-\frac{\tau_{s}^{2}}{4\mu}-\frac{\mu}{4}+\frac{\nu_{s}}{4\sqrt{2ma_{2}}}\right]R_{s}\left(\mu\right)=0.
\label{11}
\end{eqnarray}
Thus, in order to have a regular solution at the origin, we take the solution of  (\ref{11}) in the form:
\begin{eqnarray}
R_{s}\left(\mu\right)=e^{-\frac{\mu}{2}}\,\mu^{\frac{\left|\tau_{s}\right|}{2}}\,F_{s}\left(\mu\right).
\label{12}
\end{eqnarray}
Substituting (\ref{12}) into (\ref{11}), we obtain 
\begin{eqnarray}
\mu\,\frac{d^{2}F_{s}}{d\mu^{2}}+\left[\left|\tau_{s}\right|+1-\mu\right]\frac{dF_{s}}{d\mu}+\left[\frac{\nu_{s}}{4\sqrt{2ma_{2}}}-\frac{\left|\tau_{s}\right|}{2}-\frac{1}{2}\right]\,F_{s}=0,
\label{13}
\end{eqnarray}
which corresponds to the confluent hypergeometric equation or the Kummer equation \cite{abra}. The regular solution at the origin is called the Kummer function of first kind, which is given by $F_{s}\left(\mu\right)=F\left[\frac{\left|\tau_{s}\right|}{2}+\frac{1}{2}-\frac{\nu_{s}}{4\sqrt{2ma_{2}}},\left|\tau_{s}\right|+1,\mu\right]$ \cite{abra}. Hence, in order to obtain a finite solution everywhere, we must impose the condition where the confluent hypergeometric series becomes a polynomial of degree $n$ \cite{landau,abra}, where $n=0,1,2,\ldots$. This occurs when 
\begin{eqnarray}
\frac{\left|\tau_{s}\right|}{2}+\frac{1}{2}-\frac{\nu_{s}}{4\sqrt{2ma_{2}}}=-n.
\label{14}
\end{eqnarray} 

By taking the parameters defined in (\ref{10}), then, the condition (\ref{14}) yields
\begin{eqnarray}
\mathcal{E}^{2}_{n,\,l}&=&m^{2}+4\sqrt{2ma_{2}}\left[n+\frac{\left|l+\frac{1}{2}\left(1-s\right)-\frac{\phi}{\phi_{0}}+s\sqrt{2ma_{1}}\right|}{2}+s\frac{\left(l+\frac{1}{2}\left(1-s\right)-\frac{\phi}{\phi_{0}}\right)}{2}+1\right]\nonumber\\
[-2mm]\label{15}\\[-2mm]
&+&4m\sqrt{a_{1}a_{2}}.\nonumber
\end{eqnarray}

The discrete spectrum of energy (\ref{15}) corresponds to the energy levels of bound states of a spin-half charged particle confined to a two-dimensional quantum ring in a system described by the Dirac equation in the presence of a magnetic flux. We can see that the energy levels (\ref{15}) depend on the control parameters $a_{1}$ and $a_{2}$, which allow us to compare with condensed matter systems since we can identify the control parameters in such system. We should observe the flux dependence of the relativistic energy levels (\ref{15}), where there exists a periodicity for $\phi\rightarrow\phi-\phi_{0}$, that is, we have that $\mathcal{E}_{n,\,l}\left(\phi-\phi_{0}\right)=\mathcal{E}_{n,\,l+1}\left(\phi\right)$. This flux dependence of the energy levels gives rise to the arising of persistent currents \cite{by} in the two-dimensional quantum ring given by
\begin{eqnarray}
\mathcal{I}&=&-\sum_{n,\,l}\frac{\partial\mathcal{E}_{n,\,l}}{\partial\phi}\nonumber\\
&=&\frac{q}{2\pi}\sum_{n,\,l}\,\sqrt{2ma_{2}}\left(\frac{\tau_{s}}{\left|\tau_{s}\right|}+1\right)\nonumber\\
&\times&\left\{m^{2}+4\sqrt{2ma_{2}}\left[n+\frac{\left|\tau_{s}\right|}{2}+s\frac{\zeta_{s}}{2}+1\right]+4m\sqrt{a_{1}a_{2}}\right\}^{-1/2}.
\label{16}
\end{eqnarray}

We must observe that the persistent current (\ref{16}) depends on the control parameters $a_{1}$ and $a_{2}$, and the quantum numbers $n$ and $l$. Comparing with the Tan-Inkson model \cite{tan}, where the persistent currents do not depend on the quantum number $n$, we have that the present model allows us to get the information about the quantum number $n$ for persistent currents in condensed matter systems described by the Dirac equation. We should also note that the persistent current (\ref{16}) is also a periodic function of the magnetic flux $\phi$.

Now, let us make a discussion about the nonrelativistic limit of the energy levels (\ref{15}). We can obtain the nonrelativistic limit of the energy levels (\ref{15}) by applying the Taylor expansion, up to the first order terms. Thus, the Taylor expansion yields
\begin{eqnarray}
\mathcal{E}_{n,\,l}&\approx& m+\sqrt{\frac{8a_{2}}{m}}\left[n+\frac{\left|l+\frac{1}{2}\left(1-s\right)-\frac{\phi}{\phi_{0}}+s\sqrt{2ma_{1}}\right|}{2}+s\frac{\left(l+\frac{1}{2}\left(1-s\right)-\frac{\phi}{\phi_{0}}\right)}{2}+1\right]\nonumber\\
[-2mm]\label{17}\\[-2mm]
&+&2\sqrt{a_{1}a_{2}}.\nonumber
\end{eqnarray} 

Hence, we can see that the nonrelativistic energy levels (\ref{17}) recover the results of the Tan-Inkson model \cite{tan} for a spin-half charged particle confined to a two-dimensional quantum ring, whose frequency depends on the control parameter $a_{2}$, that is, $\omega=\sqrt{\frac{8a_{2}}{m}}$. We can also note the flux dependence of the nonrelativistic energy levels (\ref{17}), with periodicity being given by $\phi\rightarrow\phi-\phi_{0}$, that is, we have $\mathcal{E}_{n,\,l}\left(\phi-\phi_{0}\right)=\mathcal{E}_{n,\,l+1}\left(\phi\right)$. Thus, the persistent currents are:
\begin{eqnarray}
\mathcal{I}=-\sum_{n,\,l}\frac{\partial\mathcal{E}_{n,\,l}}{\partial\phi}\approx\frac{q}{4\pi}\sum_{l}\sqrt{\frac{8a_{2}}{m}}\,\left[\frac{\tau_{s}}{\left|\tau_{s}\right|}+1\right],
\label{18}
\end{eqnarray}
which can also be obtained from (\ref{16}) by applying the Taylor expansion, up to the terms of order $\mathcal{O}\left(m^{-2}\right)$. Note that, the persistent current (\ref{18}) depends on the control parameter $a_{1}$ and $a_{2}$, and the quantum number $l$, but not the quantum number $n$. This result agrees with the Tan-Inskon model when we consider a spin-half quantum particle. We also obverse that the persistent current (\ref{18}) is a periodic function of the magnetic flux $\phi$.

At this moment, let us come back to the minimal coupling (\ref{2}). One should note that by taking the control parameter $a_{1}=0$ in the minimal coupling (\ref{2}), we recover the Dirac oscillator \cite{do1}. For quantum systems in condensed matter described by the Dirac equation, by taking the control parameter $a_{1}=0$, the minimal coupling (\ref{2}) describes the confinement of a spin-half particle to a quantum dot. Thus, the energy levels corresponding to the confinement of a spin-half particle in a condensed matter system described by the Dirac equation are
\begin{eqnarray}
\mathcal{E}^{2}_{n,\,l}=m^{2}+4\sqrt{2ma_{2}}\left[n+\frac{\left|l+\frac{1}{2}\left(1-s\right)-\frac{\phi}{\phi_{0}}\right|}{2}+s\frac{\left(l+\frac{1}{2}\left(1-s\right)-\frac{\phi}{\phi_{0}}\right)}{2}+1\right],
\label{18a}
\end{eqnarray}   
where we can also see the flux dependence of the energy levels (\ref{18a}), that is, we also have $\mathcal{E}_{n,\,l}\left(\phi-\phi_{0}\right)=\mathcal{E}_{n,\,l+1}\left(\phi\right)$. Moreover, the persistent currents inside the quantum dot are
\begin{eqnarray}
\mathcal{I}&=&-\sum_{n,\,l}\frac{\partial\mathcal{E}_{n,\,l}}{\partial\phi}\nonumber\\
&=&\frac{q}{2\pi}\sum_{n,\,l}\,\sqrt{2ma_{2}}\left(\frac{\zeta_{s}}{\left|\zeta_{s}\right|}+1\right)\times\left\{m^{2}+4\sqrt{2ma_{2}}\left[n+\frac{\left|\zeta_{s}\right|}{2}+s\frac{\zeta_{s}}{2}+1\right]\right\}^{-1/2},
\label{18b}
\end{eqnarray}
and the nonrelativistic limit of the energy levels (\ref{18a}) can also be obtained by applying the Taylor expansion, up to the first order terms. The corresponding nonrelativistic energy levels are
\begin{eqnarray}
\mathcal{E}_{n,\,l}\approx m+\sqrt{\frac{8a_{2}}{m}}\left[n+\frac{\left|l+\frac{1}{2}\left(1-s\right)-\frac{\phi}{\phi_{0}}\right|}{2}+s\frac{\left(l+\frac{1}{2}\left(1-s\right)-\frac{\phi}{\phi_{0}}\right)}{2}+1\right],
\label{18c}
\end{eqnarray}
which recover the results obtained by Tan and Inkson in \cite{tan} for a spin-half particle. We also have a flux dependence in (\ref{18c}) with a periodicity $\mathcal{E}_{n,\,l}\left(\phi-\phi_{0}\right)=\mathcal{E}_{n,\,l+1}\left(\phi\right)$, therefore the persistent current are \cite{dot3,fur4}
\begin{eqnarray}
\mathcal{I}=-\sum_{n,\,l}\frac{\partial\mathcal{E}_{n,\,l}}{\partial\phi}\approx\frac{q}{4\pi}\sum_{l}\sqrt{\frac{8a_{2}}{m}}\,\left[\frac{\zeta_{s}}{\left|\zeta_{s}\right|}+1\right],
\label{18d}
\end{eqnarray}
which can be obtained from (\ref{18b}) by applying the Taylor expansion, up to the terms of order $\mathcal{O}\left(m^{-2}\right)$.

Furthermore, by taking the control parameter $a_{2}=0$ in (\ref{2}), we have that no bound states can be achieved anymore. This case corresponds to the interaction between the spin-half particle and a quantum antidot, where there is no confinement of the quantum particle \cite{tan}.

At this moment, let us obtain the Dirac spinors corresponding to the positive-energy solution of (\ref{3}). First of all, we should note that by taking the radial wave functions (\ref{12}), we can write (\ref{8}) in the following form
\begin{eqnarray}
\eta_{s}=e^{i\left(l+\frac{1}{2}\right)\varphi}\,e^{ikz}\,e^{\sqrt{\frac{ma_{2}}{2}}\,\rho^{2}}\,\left(2ma_{2}\right)^{\frac{\left|\tau_{s}\right|}{4}}\,\rho^{\left|\tau_{s}\right|}\,F\left[-n,\left|\tau_{s}\right|+1,\sqrt{2ma_{2}}\,\rho^{2}\right].
\label{19}
\end{eqnarray}
Thus, in order to obtain the solutions of the Dirac equation (\ref{3}), we must solve the coupled equations given in (\ref{5}) and (\ref{6}). Substituting (\ref{19}) into (\ref{6}), we can obtain the solutions for the two-spinor $\chi$. Hence, the positive-energy solutions of the Dirac equation (\ref{3}) corresponding to the parallel components to the $z$ axis are
\begin{eqnarray}
\psi_{+}&=&f_{+}\,F\left[-n,\left|\tau_{+}\right|+1,\sqrt{2ma_{2}}\,\rho^{2}\right]\left(
\begin{array}{c}
1\\
0\\
0\\
\frac{i}{\left(\mathcal{E}+m\right)}\left[2\sqrt{2ma_{2}}\,\rho+\frac{\left(\zeta_{+}-\left|\tau_{+}\right|+\sqrt{2ma_{1}}\right)}{\rho}\right]\\
\end{array}\right)\nonumber\\
&+&\frac{i\,f_{+}}{\left(\mathcal{E}+m\right)}\,\left(\frac{2n\sqrt{2ma_{2}}\,\rho}{\left|\tau_{+}\right|+1}\right)\,F\left[-n+1,\left|\tau_{+}\right|+1,\sqrt{2ma_{2}}\,\rho^{2}\right]\left(
\begin{array}{c}
0\\
0\\
0\\
1\\
\end{array}\right),
\label{20}
\end{eqnarray} 
and the antiparallel components to the $z$ axis are 
\begin{eqnarray}
\psi_{-}&=&f_{-}\,F\left[-n,\left|\tau_{-}\right|+1,\sqrt{2ma_{2}}\,\rho^{2}\right]\left(
\begin{array}{c}
0\\
1\\
\frac{i}{\left(\mathcal{E}+m\right)}\left[2\sqrt{2ma_{2}}\,\rho-\frac{\left(\zeta_{-}+\left|\tau_{-}\right|-\sqrt{2ma_{1}}\right)}{\rho}\right]\\
0\\
\end{array}\right)\nonumber\\
&+&\frac{i\,f_{-}}{\left(\mathcal{E}+m\right)}\,\left(\frac{2n\sqrt{2ma_{2}}\,\rho}{\left|\tau_{-}\right|+1}\right)\,F\left[-n+1,\left|\tau_{-}\right|+1,\sqrt{2ma_{2}}\,\rho^{2}\right]\left(
\begin{array}{c}
0\\
0\\
1\\
0\\
\end{array}\right),
\label{21}
\end{eqnarray} 
where we have defined in (\ref{20}) and (\ref{21}) the following parameters: 
\begin{eqnarray}
f_{\pm}=C\,e^{-i\mathcal{E}\,t}\,e^{i\left(l+\frac{1}{2}\right)\varphi}\,e^{ikz}\,e^{\sqrt{\frac{ma_{2}}{2}}\,\rho^{2}}\,\left(2ma_{2}\right)^{\frac{\left|\tau_{\pm}\right|}{4}}\,\rho^{\left|\tau_{\pm}\right|},
\label{22}
\end{eqnarray}
with $C$ being a constant. The spinors (\ref{20}) and (\ref{21}) correspond to the positive-energy solutions of the Dirac equation (\ref{3}). Note that the same procedure can be used in order to obtain the negative solutions of the Dirac equation (\ref{3}).

{\it In conclusion}, we have proposed a model for studying the confinement of a spin-half particle to a two-dimensional ring in quantum systems described by the Dirac equation, since these quantum systems are characterized by a linear dispersion with the velocity. This model consists in introducing a new minimal coupling into the Dirac equation which depends on two control parameters. In this way, we have shown that we can obtain a generalization of the Tan-Inkson model for a two-dimensional quantum ring \cite{tan} for systems described by the Dirac equation, where the energy levels depend on the control parameters, and in the nonrelativistic limit, we recover the nonparabolic energy levels obtained in the Tan-Inkson model \cite{tan}, but for a spin-half particle. Moreover, we have considered the presence of a magnetic flux on the $z$ axis, and shown that the energy levels are flux dependent. Thus, due to the flux dependence of the energy levels, we have discussed the arising of persistent currents in the two-dimensional quantum ring for condensed matter systems described by the Dirac equation. 

We have also shown, by taking the control parameter $a_{1}=0$, that we can recover the Dirac oscillator \cite{do1}, and study the confinement of a spin-half particle to a quantum dot in a condensed matter system described by the Dirac equation. Hence, we have obtained the energy levels, and discussed the arising of persistent currents inside the quantum dot. Moreover, we have discussed the case of the interaction between the spin-half particle with a antidot by taking the control parameter $a_{2}=0$, and shown that no bound states can be achieved. Finally, we have obtained the Dirac spinors corresponding to the positive-energy solutions.

We would like to thank CNPq (Conselho Nacional de Desenvolvimento Cient\'ifico e Tecnol\'ogico - Brazil) for financial support.

\end{document}